# Nonlinear waves, modulations and rogue waves in the modular Korteweg – de Vries equation

A.V. Slunyaev[1,2], A.V. Kokorina[1], E.N. Pelinovsky[1,2]

[1]Institute of Applied Physics of the Russian Academy of Sciences, Nizhny Novgorod, Russia
[2]National Research University Higher School of Economics, Nizhny Novgorod, Russia

**Abstract**

Effects of nonlinear dynamics of solitary waves and wave modulations within the modular (also known as quadratically cubic) Korteweg – de Vries equation are studied analytically and numerically. Large wave events can occur in the course of interaction between solitons of different signs. Stable and unstable (finite-time-lived) breathers can be generated in inelastic collisions of solitons and from perturbations of two polarities. A nonlinear evolution equation on long modulations of quasi-sinusoidal waves is derived, which is the modular or quadratically cubic nonlinear Schrödinger equation. Its solutions in the form of envelope solitons describe breathers of the modular Korteweg – de Vries equation. The instability conditions are obtained from the linear stability analysis of periodic wave perturbations. Rogue-wave-type solutions emerging due to the modulational instability in the modular Korteweg – de Vries equation are simulated numerically. They exhibit similar wave amplification, but develop faster than in the Benjamin – Feir instability described by the cubic nonlinear Schrödinger equation.

# Contents

# 1. Introduction

In this work we study nonlinear wave dynamics governed by the so-called modular Korteweg – de Vries equation in the form

$$u_t + 6|u|u_x + u_{xxx} = 0, \qquad (1)$$

which describes the real function, $u(x,t)$, of two variables denoting coordinate and time. The equation is written in dimensionless normalized form for the sake of convenience, and may be considered as a modification of the classic Korteweg – de Vries (KdV) equation with quadratic nonlinear term

$$v_t + 6vv_x + v_{xxx} = 0. \qquad (2)$$

Various generalizations of fundamental equations of mathematical physics with modified nonlinear and dispersive laws have been considered in a number of recent researches, see e.g. [Rudenko, 2013, 2016; Rosenau & Zilburg, 2018; Amiranashvili & Tobisch, 2019; Vasilieva & Rudenko, 2020; Tobisch & Pelinovsky, 2020; Pelinovsky et al., 2021a,b; Friedman et al., 2023; Garralon-Lopez et al., 2023; Pelinovsky et al., 2023] and references therein. In addition to providing a broader view on the general theory of nonlinear waves and revealing new effects of nonlinear dynamics, many of these equations find particular applications in various physical problems: as the continuous limit of anharmonic oscillators in the elasticity theory, in application to plasma and magma dynamics, surface waves on vorticity discontinuities, sedimentation and so on.

An evolution equation in the form similar to (1) was probably first written by Rudenko (2013). It was pointed in that work that the phase plane for stationary solutions of this equation looks qualitatively similar to the case of the focusing modified KdV equation (mKdV), which appears from (2) when the quadratic nonlinear term $vv_x$ is replaced by the cubic one, $v^2v_x$. Therefore, the existence of solutions with permanent shapes in the form of periodic waves and solitons was anticipated. Based on the dual similarity of the equation (1) with the quadratic and cubic KdV equations, it was named in [Rudenko, 2013] quadratically cubic KdV equation. Three conservation laws of this equation were presented in that work; the question of the integrability of the equation was raised but not answered. Meanwhile, the classic and modified KdV equations are both completely integrable by means of the Inverse Scattering Transform, see e.g. [Novikov et al., 1984; Drazin & Johnson, 1993].

The issue of well-posedness of the initial-value problem for the modular KdV equation with decaying perturbations was addressed in the recent work [Friedman et al., 2023], where this equation was considered as a member of a more general class of generalizations of the KdV equation with the absolute value incorporated into the nonlinearity characterized by different powers, including fractional. In [Friedman et al., 2023] they also considered by virtue of numerical simulations the general picture of the initial-value problem for decaying perturbations in the form of single and several humps, and also studied inelastic interactions between solitons, and found the generation of breathers in dispersive tails.

In our work we focus on the study of nonlinear wave dynamics within the modular KdV equation (1) in the context of generation of large waves; some preliminary results were announced in [Pelinovsky et al., 2022]. At first, in Sec. 2.2 we construct general solutions of the modular KdV equation in the form of waves which travel with permanent shapes and constant velocities. The classic KdV theory for cnoidal waves is used as the background for this task; it is briefly reproduced in Sec. 2.1. The analysis of the smoothness property of the obtained solutions is given in Appendix A. Solitons of different signs are the limiting cases of the constructed solutions when the wave length tends to infinity. The equation on slow modulations of quasi-sinusoidal nonlinear waves is discussed in Sec. 3 with the derivation presented in Appendix B. The linear stability analysis of perturbed periodic waves within the envelope equation is performed in Appendix C. It reveals that the waves suffer from the modulational instability, and hence rogue wave type solutions should exist in the system. The analysis is further extended in Secs. 4-5 by means of the direct numerical simulations of the modular KdV equation. Inelastic effects of soliton interactions are studied in Sec. 4, where the scenarios of large wave generation and the evidence of formation of stable and unstable breathers are discussed. The nonlinear stage of the modulational instability with generation of greatly amplified waves is considered in Sec. 5. Main findings of the research are summarized in the Conclusions.

## 2. Stationary nonlinear wave solutions

If the function $u(x,t)$ is either strictly positive or negative, the modular equation (1) reduces to the classic Korteweg – de Vries equation in the form (2) where $v(x,t) = |u(x,t)|$. The KdV equation (2) is known to possess soliton solutions

$$v_{sol}(x,t) = \frac{A}{\cosh^2\left[\sqrt{\frac{A}{2}}(x - x_0 - Vt)\right]}, \quad V = 2A > 0, \tag{3}$$

where $A > 0$ is the soliton amplitude, $V > 0$ is its velocity and $x_0$ is the reference position. The KdV equation is integrable by means of the Inverse Scattering Transform, see books [Novikov et al., 1984; Drazin & Johnson, 1993]. The solitons (3) are strictly positive, but give exact solutions to the modular KdV equation which may have either polarity: $u_{sol}(x,t) = \pm v_{sol}(x,t)$. As $N$-soliton solutions of the KdV equation $v_N(x,t)$ are strictly positive too [Gardner et al., 1974], they are exact solutions of the modular KdV equation as well, $u_N = \pm v_N(x,t)$. However, the cases when soltons of different signs coexist and interact in the modular KdV equation cannot be described in terms of the soliton solutions of the classic KdV equation. The effects of dynamics of solitons related to the specifics of the modular KdV equation were considered in [Friedman et al., 2023]; they will be further studied in Sec. 4.

Solitons of the KdV equation are the limiting case of a more general solution in form of regular waves with permanent shapes, known as cnoidal waves. In this section we first recall the main features of the cnoidal solutions of the KdV equation and then construct the analytic smooth solutions of the modular KdV equation in the form of nonlinear regular waves.

### 2.1. Cnoidal waves of the KdV equation

The stationary nonlinear wave solutions of the classic KdV equation are described by the first-order ordinary differential equation which follows from (2) after two repeating integrations:

$$\frac{1}{2}\left(\frac{dv}{d\xi}\right)^2 + P(v) = W, \quad P(v) = v^3 - \frac{V}{2}v^2 + Cv. \tag{4}$$

Here the ansatz $v(x,t) = v(\xi,t)$, $\xi = x - Vt$ is used, having the constant velocity parameter $V$ to be defined; $P(v)$ plays the role of the associated with the system mechanical potential energy; $C$ and $W$ are two constants of integration (see details in e.g. [Ostrovsky & Potapov, 1999; Pelinovsky et al., 2021b]). The cubic polynomial equation $P = W$ may have up to three real roots $v_3 \leq v_2 \leq v_1$ (see Fig. 1a), which fully determine the solution of the differential equation (4). Namely, the cnoidal wave solution requires all the roots to be different, and reads (accurate to the arbitrary shift of the coordinate $\xi$)

$$v_{cn}(\xi) = v_2 + 2a\,\text{cn}^2\left(\frac{\sqrt{a}}{s}\xi; s\right), \qquad (5)$$

$$a = (v_1 - v_2)/2 > 0, \quad V = 2(v_1 + v_2 + v_3), \quad s = \sqrt{\frac{v_1 - v_2}{v_1 - v_3}}, \quad 0 < s < 1,$$

$$\lambda = \sqrt{2}\int_{v_2}^{v_1}\frac{dr}{\sqrt{(v_1 - r)(v_2 - r)(v_3 - r)}} = 2K(s)\frac{s}{\sqrt{a}}, \qquad K(s) = \int_0^{\pi/2}\frac{d\varphi}{\sqrt{1 - s^2\sin^2\varphi}}.$$

Here cn(·;s) stands for the elliptic cosine function with the elliptic modulus $s$; $K(s)$ is the complete elliptic integral of the first kind; the parameters $a$ and $\lambda$ denote the wave amplitude and the wave length respectively. The cnoidal wave (5) is confined between two roots: $v_2 \leq v_{cn} \leq v_1$. The soliton solution (3) appears from (5) in the limit when $v_3$ and $v_2$ coalesce, what occurs when $s \to 1$ and yields $\lambda \to \infty$ (due to the infinite growth of $K$); the soliton propagates on the background level $v_b = v_2 = v_3$ with the velocity $V = 6v_b + 4a$ which is composed of the background flow speed $6v_b$ and the intrinsic speed of the soliton with the height $A = 2a$. The cnoidal waves become essentially asymmetric with respect to the horizontal line, having sharp crests and smoothed shallow troughs, when the parameter $s$ approaches one.

The solution (5) is specified by 3 free parameters (the roots $v_1$, $v_2$, $v_3$ or the parameters $V$, $C$, $W$). Under the requirement that the mean of the solution calculated for one wave period is zero, the cnoidal solution (5) may be written in the zero-mean form [Ostrovsky & Potapov, 1999]

$$v_{cn}(\xi) = 2a\left[\text{cn}^2\left(\frac{\sqrt{a}}{s}\xi; s\right) + \frac{1}{s^2}\left(1 - \frac{E(s)}{K(s)}\right) - 1\right] = \frac{2a}{s^2}\left[\text{dn}^2\left(\frac{\sqrt{a}}{s}\xi; s\right) - \frac{E(s)}{K(s)}\right], \qquad (6)$$

where $E(s)$ is the complete elliptic integral of the second kind, and only two parameters remain free. For a given wavenumber $k = 2\pi/\lambda$ and amplitude $a$, the frequency $\omega = Vk$ of the regular wave (6) reads

$$\omega(k,a) = -k^3\left(\frac{2K(s)}{\pi}\right)^2\left(3\frac{E(s)}{K(s)} + s^2 - 2\right), \quad \text{where} \quad \frac{a}{k^2} = \left(\frac{sK(s)}{\pi}\right)^2. \qquad (7)$$

The amplitude enters the nonlinear dispersion relation $\omega(k,a)$ through the parameter $s$ of the elliptic integral.

The balance between effects of nonlinearity and dispersion in Eq. (2) may be evaluated through the similarity parameter which for the characteristic wave amplitude $a$ and the wavenumber $k$ is proportional to the combination $a/k^2$. In the shallow water theory it is known as the Ursell parameter $Ur$, which will be determined in this work as follows,

$$Ur = \frac{a}{k^2}. \qquad (8)$$

Indeed, according to (7), the product $sK(s)$, which specifies the elliptic integral, depends on the combination $a/k^2$. Alternatively, the elliptic integral parameter $s$ may be used to characterize the balance between wave nonlinearity and dispersion of the periodic wave solution too. In the small-amplitude limit $s \to 0$ (and then $K \approx \pi/2(1 + s^2/4)$, $E \approx \pi/2(1 - s^2/4)$, $Ur \approx s^2/4$) the nonlinear dispersion relation (7) reduces to the linear dispersion relation $\omega_{lin}$ with small amplitude correction:

$$\omega(k)\xrightarrow[s\to 0]{}\omega_{lin}\left(1+3\left(\frac{a}{k^2}\right)^2+O(s^6)\right), \qquad \omega_{lin}(k)=-k^3. \tag{9}$$

In the soliton limit $s \to 1$ we have $E \to 1$, and $K$ grows infinitely. Then the relations (7) yield the linear relation between $\omega$ and $k$:

$$\omega(k,a)\xrightarrow[s\to 1]{}4ka, \tag{10}$$

what gives the speed of a soliton (3) with the height $A = 2a$: $V = \omega/k = 4a$.

The asymptotic frequencies (9) and (10) are of different signs giving that small-amplitude cnoidal waves travel to the left, whereas waves characterized by sufficiently large similarity parameter $s$ propagate in the direction of the coordinate axis. The separating value of the parameter $s$, which corresponds to waves that do not move in the reference of Eq. (2), is specified by the condition

$$3\frac{E(s)}{K(s)}+s^2-2=0, \tag{11}$$

which is satisfied when $s \approx 0.98$.

**2.2. Stationary periodic waves of the modular KdV equation**

Given that any positive solution of the KdV equation (2) taken with the same or negative sign will be a solution of the modular KdV equation (1), one can construct sign-changing stationary solutions of the modular KdV equation from sign-preserving intervals of the cnoidal solutions (5). Smoothness conditions should apply in the linkage points where the stationary wave $u(\xi)$, $\xi = x - Vt$, changes the sign with the purpose to make the obtained solution at least three times differentiable, so that it may be called the solution of the Eq. (1) in the strict sense. It may be shown (see details in Appendix A) that there are only two ways to construct smooth sign-changing stationary solutions of the modular KdV equation, which correspond to the cuts of the original potential $P(v)$ (see Eq. (4)) at the points of the local minimum or the local maximum as shown in Fig. 1b and Fig. 1c respectively. These solutions correspond to the following choices of the parameters in $P(v)$: $C = 0$ and $V < 0$ (Fig. 1b), and $C = 0$ and $V > 0$ (Fig. 1c). The original KdV potential $P(v)$ and the potential for the modular KdV equation $\Pi(u)$ are given in the figures with solid blue and dashed red curves respectively. As shown in Appendix A, these solutions are at least three times differentiable and correspond to permanent waveforms traveling with the constant velocities $V$. For any amplitude, they are symmetric with respect to the zero level $u = 0$, in contrast to the cnoidal waves of the quadratic KdV equation.

<u>Periodic waves with negative velocities $V < 0$</u>

In the case shown in Fig. 1b the wave velocity is negative or zero, $V \leq 0$, the condition $P(v) = W$ possesses three different real roots. The distance between two consecutive zero crossings by the KdV cnoidal wave solution (5) $v(\xi) = 0$ corresponds to the half length of the modular KdV solution, therefore the full wave length is determined by its doubled value:

$$\lambda=\frac{4\sqrt{2}s}{\sqrt{v_1-v_2}}F(\varphi_*,s), \quad F(\varphi_*,s)=\int_0^{\varphi_*}\frac{d\varphi}{\sqrt{1-s^2\sin^2\varphi}}, \quad \varphi_*=\arccos\left(\sqrt{\frac{v_2}{v_2-v_1}}\right), \tag{12}$$

where $F(\cdot;s)$ is the incomplete elliptic integral of the first kind, and the argument $\varphi_*$ corresponds to the point where the solution $v(\xi)$ crosses zero. Note that this point does not correspond to the mean level of the 'mother' KdV cnoidal wave solution. The maximum elevation position in the cnoidal wave is specified by the condition $\varphi = 0$. The amplitude $a$ of the wave in the modular KdV equation is determined by $a = v_1$, what may be presented in the form

$$a = (v_1 - v_2)\sin^2 \varphi_*. \qquad (13)$$

Recall that in (12) and (13) the value $(v_1 - v_2)/2$ is the amplitude of the 'mother' KdV cnoidal wave (5).

One period $-\lambda/2 \leq \xi < \lambda/2$ of the nonlinear wave solution of the modular equation $u(\xi)$ can be constructed from the positive part of the KdV cnoidal function $v_{cn}(\xi)$ (5) by its continuation when taken with different sign, as follows:

$$u(\xi) = \begin{cases} -v_{cn}\left(\xi + \dfrac{\lambda}{2}\right), & -\dfrac{\lambda}{2} \leq \xi < -\dfrac{\lambda}{4} \\ v_{cn}(\xi), & -\dfrac{\lambda}{4} \leq \xi \leq \dfrac{\lambda}{4} \\ -v_{cn}\left(\xi - \dfrac{\lambda}{2}\right), & \dfrac{\lambda}{4} \leq \xi < \dfrac{\lambda}{2} \end{cases}, \qquad (14)$$

where $\lambda$ is specified by (12). The solution in infinite line is obtained by periodic continuation of the wave (14).

Periodic waves with positive velocities $V > 0$

The situation shown in Fig. 1c is characterized by positive velocities $V > 0$ and the integration parameters $C = 0$, $W \geq 0$, so that the condition $P(v) = W$ yields only one real root when $W > 0$ and two roots in the particular case $W = 0$. It is obvious that in the latter case $W = 0$ the solutions coincide with the KdV solitons on a zero background (taken with the sign plus or minus), given in Eq. (3). From the condition on the wave amplitude $a = v_1 > 0$ in form $P(a) = a^3 - V/2\, a^2 = W \geq 0$ we obtain the upper limit of the wave velocity, $V \leq V_s$, $V_s = 2a$, which corresponds to the soliton solution (3) with the height $A = a$.

The solution which corresponds to the range of velocities $0 < V < V_s$ and $W > 0$ is described by the integral

$$\pm \xi = \int_0^u \frac{dr}{\sqrt{2(\Pi(a) - \Pi(r))}}, \qquad \Pi(u) = |u|u^2 - \frac{V}{2}u^2, \qquad (15)$$

which is obtained from the first-order ordinary differential equation (4) taking into account the extension of the potential energy function to negative values of $u$ (alternatively, see (A.3) for $W_+ = W_- = W$). In (15) an arbitrary shift of the coordinate $\xi$ can be applied. This integral may be formally written in terms of the elliptic integral with a complex parameter, but for the needs of the present work it is more constructive to calculate and analyze it numerically. The denominator in the integral (15) tends to zero when $r$ approaches $a$, what complicates the numerical calculation of the integral. For the purpose of numerical integration it is more advantageous to represent (15) in the form:

$$\pm \sqrt{2}\xi = \int_0^u \left[\frac{1}{\sqrt{Q(r)}} - \frac{1}{\sqrt{Q(a)}}\right] \frac{dr}{\sqrt{(a-r)}} + 2\frac{\sqrt{a} - \sqrt{u}}{\sqrt{Q(a)}}, \qquad (16)$$

$$Q(u) = u^2 + \left(a - \frac{V}{2}\right)u + \frac{\Pi(a)}{a}, \qquad 0 \leq u \leq a.$$

The solution (16) is formally identical to (15) when $u \geq 0$ and may be used to build the wave in the interval $-\lambda/4 \leq \xi < \lambda/4$. The solution for the other half period which corresponds to the negative part of the wave, $-a \leq u < 0$, is obtained from (16) by continuation with the negative sign (similar to Eq. (14)). The total wave length $\lambda$ is determined by the relation which straightforwardly follows from (15)

$$\lambda = 4 \int_0^a \frac{dr}{\sqrt{2(\Pi(a) - \Pi(r))}}. \tag{17}$$

It was calculated numerically using the representation of the integral similar to the one in (16).

To conclude, the stationary periodic nonlinear wave solutions of the modular KdV equation are constructed, which are characterized by the amplitude $a > 0$, length $\lambda > 0$ and velocity $V < V_s$, $V_s > 0$. Any pair of these parameters may be considered as the arbitrary parameters of the solution. Then, the wavenumber $k = 2\pi/\lambda$ and the wave frequency $\omega = Vk$ follow. Few shapes of the periodic solutions of the modular KdV equation are shown in Fig. 2. It is obvious, that the similarity parameter $Ur$ in form (8) applies to the modular KdV equation (1) too; it specifies the degree of nonlinearity with respect to the wave dispersion. Waves with small $Ur$ are close to sinusoidal, while they resemble a sequence of solitons with alternating signs in the limit of relatively large nonlinearity. For the waves in Fig. 2 with the amplitude $a = 2$ the limiting velocity of a soliton is given by $V_s = 4$.

The maximal velocity $V_s = 2a$ of the modular wave is twice smaller than the speed of the KdV cnoidal wave in the soliton limit (10). In this limit the KdV cnoidal wave (6) becomes strongly asymmetric, whereas the modular nonlinear waves are always symmetric. The comparison between celerities of nonlinear waves of the two equations will be more consistent when the crest amplitude $a_{cr}$ (i.e., the maximum of the wave with respect to the mean) is used to characterize the wave intensity. Then the amplitudes $a_{cr}$ of the KdV cnoidal wave (6) and of the modular KdV solution coincide in the both limits: when the Usell number (8) is either small or large. Due to the symmetry of waves of the modular equation their amplitudes $a$ are identical to $a_{cr}$. For the KdV cnoidal waves on zero background (6) their amplitudes $a_{cn} = (v_1 - v_2)/2$ and crest amplitudes $a_{cr}$ are related as follows:

$$a_{cr} = \frac{a_{cn}}{s^2}\left(1 - \frac{E(s)}{K(s)}\right). \tag{18}$$

The nonlinear dispersion relation for the modular equation is compared with the one for the classic KdV equation in Fig. 3. The dependence for the modular equation was also verified in the direct numerical simulation. The curves in the figure are plotted for the two models with the same values of the wave crest amplitude, $a_{cr}$. Note that the curves in Fig. 3 are continuous smooth functions everywhere including the points $V = 0$, which divide the two options of constructing the alternating stationary solutions of the modular equation, described above. In Fig. 3, the condition $V = 0$ corresponds to the locus where $\omega = 0$, given by the boundary between shaded and clear areas. Thus, the solutions for negative and positive velocities smoothly transform one into the other. Note that according to Fig. 2 the nonlinear waves with $V = 0$ are still very close to sinusoidal.

As one can see from Fig. 3, the dependences for the classic and modular KdV equations have the same asymptotics when $k$ tends to zero, which correspond to the soliton limit. In this limit the wave phase velocity approaches the speed of a soliton $V = 2A$ with the amplitude $A = a_{cr}$: $\omega/k - \omega_{lin}/k \approx \omega/k \approx 2a_{cr}$ when $k \to 0$ (see Eqs. (3) and (10)). The straight dashed lines, which correspond to the soliton velocities $(\omega - \omega_{lin})/k = 2a_{cr}$, separate the dependences for the classic and the modular KdV equations (below and above these lines, respectively). Hence, frequencies of the modular KdV solutions (thick curves) exceed those of the quadratic KdV equation (thin curves of the same color) for the same choices of wave lengths and crest amplitudes. When the amplitude $a_{cr}$ is fixed but the wavenumber grows, in the case of the quadratic KdV equation the nonlinear frequency shift $\omega_{nl} = \omega - \omega_{lin}$ decays to zero as $k^{-1}$ according to (9); within the modular equation it grows proportionally to $k$, with the proportionality coefficient $(\omega - \omega_{lin})/k = 8a_{cr}/\pi$ exceeding the value characteristic for the soliton limit (see the derivation of this estimate later in Sec. 3).

## 3. Nonlinear equation for wave modulations

As will be discussed in Secs. 4,5 nontrivial nonlinear dynamics of wave groups may be observed within the modular KdV equation, which deserves understanding. An equation for modulations of nonlinear waves of the modular equation has been derived to this end. The absolute value function in equation (1) prevents developing the traditional asymptotic theory for KdV-like equations similar to as in [Grimshaw et al., 2001; Slunyaev, 2005]. In this work we develop a modified approach described in Appendix B, which is conceptually close to the paper by [Tobisch & Pelinovsky, 2019]. As a result, the equation on the wave modulations has the form of the nonlinear Schrödinger (NLS) equation

$$i(\psi_t + c_{gr}\psi_x) + p|\psi|\psi + q\psi_{xx} = 0, \qquad (19)$$

$$c_{gr} = -3k_0^2, \qquad p = \frac{8}{\pi}k_0, \qquad q = 3k_0,$$

which describes the complex envelope $\psi(x,t)$ related to the original function $u(x,t)$ according to the formula

$$u(x,t) = \frac{1}{2}\psi(x,t)\exp(i\omega_0 t - ik_0 x) + c.c.. \qquad (20)$$

Here $k_0$ is the carrier wave number, and $\omega_0 = \omega_{lin}(k_0)$ as per (9). The derivation of (19) implies that the wave amplitude is small in the sense that waves are close to sinusoidal (the parameter $Ur = a_0/k_0^2$ should be small to some extent), and that the modulations are long compared to the carrier wave length.

The equation for modulations (19) inherits the non-analytic nonlinearity of the parent equation (1). It is formally second order in nonlinearity, but possesses the property of symmetry with respect to the sign of perturbations. Following [Rudenko, 2013], it may be called quadratically cubic NLS equation. Equation in the form (19) with two spatial coordinates was introduced in [Rudenko, 2013], where two conservation laws of this equation, and also reductions to other forms of the equation were presented.

The modular NLS equation (19) has the solution in the form of a plane wave with the amplitude $a_0 > 0$:

$$\psi_{pw}(x,t) = a_0\exp(i\omega_{nl}t), \qquad \omega_{nl} = pa_0 = \frac{8}{\pi}k_0 a_0. \qquad (21)$$

As before, the quantity $\omega_{nl}$ has the meaning of the frequency shift with respect to the carrier wave frequency $\omega_0$ due to the nonlinearity. It is proportional to the scaled wave amplitude $k_0 a_0 = 2\pi a_0/\lambda_0$ called wave steepness.

Under the condition $pq > 0$, which holds by Eq. (19), the modular NLS equation possesses an envelope soliton solution in the form

$$\psi_{es}(x,t) = \frac{B\exp\left(i\frac{2p}{3}Bt\right)}{\cosh^2\left[\sqrt{\frac{pB}{6q}}(x-x_0)\right]} = \frac{B\exp\left(i\frac{16}{3\pi}k_0 Bt\right)}{\cosh^2\left[\frac{2}{3}\sqrt{\frac{B}{\pi}}(x-x_0)\right]} \quad (22)$$

with the amplitude $B > 0$ which may be assumed positive with no loss of generality. The shape of the envelope solitons (22) sech$^2$ coincides with the shape of the classic KdV solitons (3), but the envelope solitons are approximately twice wider than solitons (3) of the same amplitude, $A = B$. The analytic solutions (21) and (22) may be further generalized by addition of arbitrary phase constants, and using the Galilean transformation.

The existence of bright envelope solitons is deeply related to the effect of the modulational instability (also called the Benjamin – Feir instability). Indeed, as follows from the analysis of linear stability of the plane wave (21) with respect to long perturbations (see details in Appendix C), it is modulationally unstable for perturbations characterized by sufficiently small wavenumbers $K$ with the growth rate $\sigma$:

$$|K| < K_{BF}, \qquad K_{BF} = 2\sqrt{\frac{2a_0}{3\pi}}, \qquad \sigma = 3|k_0 K|\sqrt{\frac{8}{3\pi}a_0 - K^2}. \quad (23)$$

The maximum growth rate of the instability and the corresponding perturbation wavenumber read

$$\sigma_{\max} = \frac{4}{\pi}k_0 a_0, \qquad K_{\max} = \frac{K_{BF}}{\sqrt{2}}. \quad (24)$$

Thus, though cnoidal waves of the classic KdV equation are known to be modulationally stable, weakly nonlinear waves within the modular equation (1) are unstable with respect to long perturbations. Note that the wave dynamics within the modular NLS equation including the instability conditions (23) are controlled by the similarity parameter $a/K^2$ which characterizes the ratio of nonlinear versus dispersive term in Eq. (19). It has the form similar to the Ursell number (8), but involves the length scale of the perturbation rather than the carrier wave length.

In Fig. 4 the nonlinear correction to a regular wave frequency described by the approximate NLS equation (19) is compared in normalized axes with the nonlinear frequency for exact periodic solutions of the modular equation (already illustrated in Fig. 3). The relative value of the nonlinear frequency shift of a periodic wave $(\omega - \omega_0)/\omega_0$, where $\omega_0 = \omega_{lin}(k_0)$ is specified in (9), gives $(\omega - \omega_0)/\omega_0 \propto \omega_{nl}/k_0^3$. Then, the relative nonlinear frequency of the plane wave solution (21) is $\omega_{nl}/k^3 \propto Ur$, where the parameter $Ur = a_0/k_0^2$ (as per (8)) is the measure of nonlinearity of the modular KdV equation solutions and, simultaneously, is the control parameter of applicability of the modular NLS equation. One may see from Fig. 4, that the solution (21) shown with the straight red solid line and the solution for exact regular waves (circles) exhibit fine agreement if the Ursell number is not large (see the inset for the interval $a_0/k_0^2 \leq 1$). The solution (21) gives the proportionality coefficient $8/\pi \approx 2.55$ between $\omega_{nl}/k^3$ and $Ur = a_0/k_0^2$ which describes the range of small $Ur$ in Fig. 4; it also describes the limit of large wavenumbers in Fig. 3 (see thick solid curves). For large parameters $a_0/k_0^2$ the wave differs from a sinusoid (see examples in Fig. 2), and the discrepancy between the two frequencies in Fig. 4 becomes obvious (but still not large in value). The modular NLS equation overestimates the nonlinear frequency. The nonlinear frequency within the modular equation for large $a_0/k_0^2$ agrees very well with the asymptotic $2a_0/k_0^2$ (the dashed line) which corresponds to the soliton limit with velocity $V_s$ being twice the soliton amplitude $A = a_0$, what is in agreement with Fig. 3.

## 4. Numerical simulation of solitons and breathers

In this study the direct numerical simulation of the modular KdV equation was performed using a pseudo-spectral scheme with three-layer finite difference discretization of time, similar to the one described in our work [Didenkulova et al., 2019]. The employed Crank-Nicolson scheme is unconditionally stable with respect to the time discretization; in the performed simulations the allowable relative error in the conservation of the integral $_{-\infty}\int^{\infty} u^2\, dx$ was of the order $10^{-5}$. Similar codes were used to perform comparative simulations within the classic KdV equation (2), and the modified KdV equation with cubic nonlinearity, when the term $vv_x$ in Eq. (2) is replaced with $v^2 v_x$. Periodic boundary conditions are imposed.

The scenario of elastic interaction of solitons well-known within the classic KdV equation is perfectly reproduced within the modular equation if the initial condition for the simulations is taken in the form of well separated solitons (3) with the same sign, see Fig. 5a. An exchange interaction between two solitons with amplitudes $A_1 = 2$ and $A_2 = 1.8$ is shown in the space-time diagram, where the horizontal axis corresponds to the reference moving with the speed of the largest soliton $V_1$. Due to the periodic boundary conditions, the soliton interactions repeat again in a completely identical way. If amplitudes of the solitons are sufficiently different in value but have the same signs, they interact following the overtaking scenario. When interact, solitons acquire different phase shifts: the faster soliton is forwarded in the direction of propagation, whereas the other soliton shifts back. This picture of phase shifts is universal for integrable KdV equations and does not depend on the soliton signs, which may be different in the case of the modified KdV equation. As was discussed above, $N$-soliton solutions of the KdV equation taken with the same or opposite sign are solutions of the modular KdV equation; the direct numerical simulation confirms that they do not exhibit structural instability.

The simulation of a similar problem but when the second soliton is taken with different sign is shown in Fig. 5b (the positive soliton is reflected by the bright trace, whereas the dark trace corresponds to the negative soliton). The solitons have the same velocities as before and collide at first at the same time, but the interaction occurs very different to the previous case: it is now clearly not elastic, see distinguishable small-amplitude waves radiated to the left (backwards) in Fig. 5b. The solitons survive the collision, but only approximately recover their shapes, what is clearly seen from the altered velocities: the highest positive soliton becomes faster (and hence even bigger), while the negative soliton slows down and reduces its height. A deeper investigation reveals that the solitons are both shifted backwards when collide. The evolutions of the wave maximum and minimum calculated in the simulation domain are shown in Fig. 6. One can see that after the first collision the soliton amplitudes are indeed not restored in full, the discrimination between soliton heights increases. Fluctuations of the curves between the consecutive events of collisions reflect oscillations of the solitons on the background of small-amplitude waves. The solitons in Fig. 5b and Fig. 6 collide six times, thanks to the periodic boundary conditions. Though the interactions are clearly not elastic, and the picture of collisions is not fully repeating, one may conclude that the solitary waves within the modular equation are rather stable.

It is found that when solitons of different signs interact, for short instants of time they form larger waves with peak amplitudes somewhat smaller than the sum of the soliton heights $|A_1| + |A_2|$ (see Fig. 6). In the integrable mKdV equation the peak value would be exactly the sum [Slunyaev & Pelinovsky, 2016]. The ratios of the wave maxima, which are actually observed in the series of numerical simulations of pairs of interacting solitons with different signs, with respect to the ideal case $|A_1| + |A_2|$ are plotted in Fig. 7. The most interesting case is when the solitons have amplitudes close by the absolute value. Within the integrable theory the amplification would be about 2. Within the modular KdV the collision is accompanied by rather strong dispersive wave radiation. As a result, the collision between solitons with amplitudes $A_1 = 2$ and $A_2 = -1.8$ leads to about 16% smaller wave amplification than the ideal $(|A_1| + |A_2|)/A_1 = 1.9$ times. Solitons with very different amplitudes have greater relative velocity and radiate noticeable weaker, but generate less amplified waves.

As shown in [Slunyaev & Pelinovsky, 2016], synchronous collisions of solitons with alternating signs lead to formation of very large waves, so-called rogue waves. The nonlinear phase shifts play an essential role in such soliton focusing. As discussed above, the soliton phase shifts in elastic collisions (between solitons of one sign) and inelastic collisions (between solitons of different signs) are different within the modular KdV equation. Therefore, there is no analytic solution to initiate the synchronous collision in the modular KdV equation between solitons with alternating signs. We arranged such interactions in sets of 3 and 4 solitons in a series of trials, and obtained waves noticeably exceeding the initial condition in amplitude. Inelastic effects significantly reduce the achieved maximum wave height compared to the ultimate limit according to the integrable frameworks, which is just the sum of heights of the interacting solitons, $|A_1| + |A_2| + |A_3| + \ldots$. The collision picture is extremely sensitive with respect to the initial locations of solitons. The wave maxima as functions of time in the series of numerical simulations of three colliding solitons with alternating signs and slightly varied initial position of the smallest soliton, are shown in Fig. 8. In a similar interaction within the integrable mKdV equation the maximum wave would have the amplitude 1.8, while it is at most 1.5 in the shown example. Two experiments from the series are also shown in Fig. 9.

The case in Fig. 9a resembles the interaction scenario within the modified KdV equation with occurrence of a large wave (see in [Slunyaev & Pelinovsky, 2016]) except for emitting small-amplitude waves. An essentially new effect is observed in the numerical experiment shown in Fig. 9b, where the largest positive soliton (it is bright and originally is standing in the reference of the plot) and the negative soliton remain for some time coupled after the triple collision, and then interact again and separate. In the course of the interaction events the order of solitons in the train changes from the sequence 'positive-negative-positive' to 'negative-positive-positive'. The process of coupling between solitons of different signs is known for the mKdV equation [Ivanychev & Fraiman, 1997]; it may occur due to perturbations which break integrability (such as numerical viscosity, variable coefficients, etc.) and correspond to the bifurcation of the spectrum of the associated scattering problem and emergence of new solutions of the integrable mKdV equation – breathers [Pelinovsky & Grimshaw, 1997].

For equations integrable by means of the Inverse Scattering Technique, the rigorous analytical analysis of the initial-value problem is possible. In particular, it was shown in [Clarke et al., 2000] that breathers of the mKdV equation are likely to appear from localized disturbances with both polarities. Indeed, this configuration of the initial condition is found to be prone to the generation of breathers within the modular KdV equation as well. A comparison of the transformation of the initial condition in the form of two adjacent boxes with different signs within the framework of the integrable mKdV equation and within the modular KdV equation is given in Fig. 10. (The shape of the initial condition was smoothed for better stability of the numerical simulation.) Within the mKdV equation (Fig. 10a) one can distinguish two solitons of different signs which propagate to the right with very close speeds faster than all other waves. Two breathers propagate to the right with smaller velocities; the slower breather oscillates more frequently. A small-amplitude dispersive wave train propagates to the left.

On the first glance, the picture of the wave evolution within the modular KdV equation in Fig. 10b looks similar; three breather-like structures propagating to the right take most of the wave energy. However, after few oscillations the solitary waves of different polarities decouple, similar to the dynamics in Fig. 9b. New portions of small-amplitude waves are radiated to the left at every event of 'turnabout' of the coupled solitary waves. To the best of our knowledge, the 'decoupling' of breathers have not been reported in the literature before. According to the performed numerical simulations, breathers of the modular KdV equation seem to be more stable when composed by slower solitons.

An example of a stable breather is also present in Fig. 10b, see the wave group indicated by an arrow, which emerges at the early stage, gradually propagates to the left and then re-appears from the right side due to the periodic boundary conditions. Similar breathers were reported in the numerical simulation of the modular KdV by [Friedman et al., 2023], see their Figure 42. The breather in Fig. 10b recovers after a series of collisions with other waves though slightly changes its speed. This localized nonlinear group is compared in Fig. 11 against the envelope soliton solution (22) & (20) of the equation on wave modulations (19). The carrier wavenumber, and also the envelope soliton amplitude and the constant phase were tuned to obtain the best fit with the group. One may conclude about very good agreement, so the group in Fig. 10b may be treated as the envelope soliton of the equation for modulations. We have checked in dedicated numerical simulations that the initial condition to the modular KdV equation taken in the form of the envelope soliton (22) with the use of (20) indeed gives life to a stable nonlinear wave group.

In Figs. 10, 11 the stable breather propagates to the left similar to dispersive waves, what reveals its relatively small degree of nonlinearity. Stable breathers which travel with positive velocity may be generated too. Such an example is given in Fig. 12, where an intense breather emerges from the initial condition in the form of two adjacent boxes of different polarities. A continuous radiation of small-amplitude waves by the breather is readily seen. Due to this effect the breather is slowly decreasing in amplitude, but so far remains stable.

## 5. Numerical simulation of the modulational instability and rogue wave generation

Instability of nonlinear waves with respect to long perturbations is known to be a regular mechanism of large-wave generation, known as rogue or freak waves [Onorato et al., 2001, 2013]. It has been stated in [Slunyaev et al., 2023] based on a number of examples from integrable systems that rogue wave solutions are inseparably linked to the modulational instability. Similar to the geophysical counterpart, the rogue wave solutions of partial differential equation are those over some background, which "appear from nowhere and disappear without a trace" [Akhmediev et al., 2009]. In this section we investigate the picture of modular rogue waves, which occur as a result of the modulational instability discovered and quantified in Sec. 3.

Within the integrable cubic Schrödinger equation the nonlinear stage of the modulational (also known as Benjamin – Feir) instability can be described in terms of exact breather solutions, determined by the spectrum of the associated scattering problem, see e.g. [Osborne, 2010]. In particular, the maximum wave amplification may be predicted based on the initial background and the perturbation characteristics. In this work, we draw the picture of the modulational instability within the modular KdV equation by virtue of the direct numerical simulation. We employ a traditional problem setup for such an analysis. The initial condition is taken in the form of a sequence of periodic waves, which are weakly perturbed in amplitude with a given perturbation length:

$$u(x, t = 0) = \left(1 + \delta \cos\left(\frac{2\pi}{L} x\right)\right) u_0(x), \quad L = \lambda_0 N_w . \qquad (25)$$

Here $u_0(x)$ is the periodic nonlinear wave solution for the modular KdV equation with the amplitude $a_0$ and length $\lambda_0$; $\delta = 0.05$ is the perturbation amplitude. The size of the periodic computation domain $L = \lambda_0 N_w$ coincides with the length of the initial perturbation, where integer $N_w$ is the number of wave periods per modulation length. In the examples in Fig. 13 the maximum waves which occur as a result of the modulational instability are shown for simulations with the wave parameters $a_0 = 0.5$, $\lambda_0 \approx 5.9$, and two perturbation lengths of $N_w = 4$ (Fig. 13a) and $N_w = 15$ (Fig. 13b) waves. Both the perturbation lengths correspond to modulationally unstable conditions according to the criterion (23). The first case is characterized by larger growth rate $\sigma$ than the second case, but the eventual amplitude of the emerged rogue wave is noticeably bigger in the second case, where it reaches the value about thrice the initial wave amplitude. Recall that the celebrated Peregrine breather within the nonlinear Schrödinger equation is characterized by the maximum amplification which is exactly 3 [Peregrine, 1983]; it is formed due to the development of infinitely long perturbation for, formally, infinitely long time. Thus, its counterpart within the modular KdV equation excited by very long perturbations is characterized by a similar amplification limit.

Quasi repetition of the modulation-demodulation processes, known as the Fermi-Pasta-Ulam-Tsingou phenomenon, was observed in numerical simulations of modulationally unstable waves within the modular KdV equation.

The evolution of the initial condition taken in the form of a sinusoidal wave with the amplitude $a$ and wavenumber $k$, such that the Ursell parameter $a/k^2$ is not small, is shown in Fig. 14. When the perturbation evolves, soliton-type waveforms appear for some instants (see Fig. 14a, the shape for $t = 0.6$) similar to the nonlinear wave shape shown in Fig. 2 with the blue curve. Thus, the nonlinear wave solutions constructed in Sec. 2 exhibit attracting property for periodic initial conditions of a general form. Note that the initial condition parameters $a$ and $k$ formally satisfy the instability criterion (23), though the wave remains globally stable in time as shown in Fig. 14b.

## 6. Conclusions

In this paper, we have constructed and studied analytically and numerically the key nonlinear solutions of the modular KdV equation, which also may be referred to as quadratically cubic KdV equation [Rudenko, 2013]. When the dynamics of solitons of one sign are concerned, this unique system possesses features of integrable soliton-generating systems, such as fully elastic collisions between solitons and existence of exact $N$-soliton solutions, and also the infinite number of conserved quantities. However, the dynamics changes drastically as soon as the perturbation has two polarities (sign-changing): solitary waves of different polarities interact inelastically; the exact periodic wave is finitely differentiable, what should apparently mean that the number of conserved laws of the modular KdV equation is finite.

Exact solutions of the modular KdV equation in the form of nonlinear periodic waves are constructed, which are differentiable everywhere at least thrice. For small amplitudes, the waves are close to sinusoidal. Waves with larger amplitudes are locally identical to crests of cnoidal wave solutions within the quadratic KdV equation, but are symmetric with respect to the horizontal line. Similar to the KdV equation, the degree of wave nonlinearity is controlled by the same similarity parameter known as the Ursell number, proportional to $a/k^2$, where $a$ is the wave amplitude and $k$ is the wavenumber. The nonlinear frequency shift of modular waves is larger in magnitude than that of the KdV cnoidal waves of similar crest amplitude. In the limit of very long waves with finite amplitude the solution tends to a sequence of classic KdV solitons with alternating signs.

The nonlinear wave dynamics within the modular KdV equation is rich; it combines effects known for the integrable KdV and mKdV equations as well as new effects. Due to inelasticity of the interaction of solitons with different signs, the efficiency of high wave generation in synchronous collisions of alternating solitons (see [Slunyaev & Pelinovsky, 2016]) is noticeably reduced. After a collision of two solitons of different signs the difference in their velocities (and in absolute values of amplitudes) increases; they both experience backward shifts when interact.

As we mentioned in [Pelinovsky et al., 2022], bound structures may be born in interactions between solitons of different signs and from perturbations of two signs; they resemble breathers of the modified KdV equation. These 'modular breathers' may be either stable or unstable. Stable breathers have the form of wave groups; these waves were observed in the numerical simulations by Friedman et al. (2023). Unstable breather-type bound structures consist of a couple of soliton-like waves of different signs. They propagate for some time as a whole exhibiting quasi-periodic exchange of the solitons' mutual positions, and eventually split into two solitons of different signs. As a result, it becomes hardly possible to predict the sign of the leading soliton which will be generated after a collision of two solitons with different signs but close velocities. Though a bifurcation of two solitons of different signs to a breather is a known effect within systems close to integrable [Ivanychev & Fraiman, 1997; Pelinovsky & Grimshaw, 1997], the reverse process has apparently not been reported by now.

The nonlinear equation for long modulations of quasi-sinusoidal waves within the modular KdV equation is derived. It has the form of the nonlinear parabolic equation (NLS equation) with the absolute value term, which is formally quadratic in nonlinearity. This modular nonlinear Schrödinger equation may be also called quadratically cubic NLS equation, following [Rudenko, 2013]. It possesses exact solutions in the form of envelope solitons, which describe very well the stable breathers observed in the numerical simulations of the modular kdV equation. Solitons of the modular KdV equation and envelope solitons of the modular envelope equation have the same $sech^2$ shape, but differ by about a factor of two in width.

Waves within the modular envelope equation are shown to be modulationally unstable. Solutions of the rogue wave type are revealed by virtue of the numerical simulation of the modular KdV equation. They correspond to growing long perturbations of regular waves, which appear almost from nowhere and disappear with almost no trace. Their main features resemble breather solutions of the integrable cubic Schrödinger equation: perturbations with larger lengths develop slower, but eventually lead to greater wave amplification, which is limited by approximately the value of 3, similar to the celebrated Peregrine breather. The modulational instability within the modular envelope equation is supported by three wave interactions, what is a faster process than the Benjamin – Feir modulational instability caused by quasi-resonant dynamics in wave quartets. Respectively, the instability conditions within the modular NLS equation are controlled by its similarity parameter, which has the form of the Ursell number $a/K^2$, but instead of the dominant wavenumber includes the perturbation wavenumber $K$ (the analogue of the Benjamin – Feir Index).

## Acknowledgements

The research in Sec. 3 was supported by the Foundation for the Advancement of Theoretical Physics and Mathematics "BASIS" No. 22-1-2-42 and by the state contract No 0030-2021-0007. The remaining sections were supported by the Russian Science Foundation Grant No. 19-12-00253.

# Appendix A. Smoothness of the stationary nonlinear waves within the modular KdV equation

Consider waves of permanent shape which travel with the constant velocity $V$: $u(x,t) = u(\xi)$, $\xi = x - Vt$. Then the modular KdV equation in partial derivatives (1) reduces to the ordinary differential equation

$$-Vu_\xi + 6|u|u_\xi + u_{\xi\xi\xi} = 0. \tag{A.1}$$

Let us analyze the equation (A.1) for the domains $u > 0$ and $u < 0$ independently, and then link the parts of the solution in the points $\xi = \xi_0$ where $u(\xi_0) = 0$, bearing in mind that the solution $u(\xi)$ should be at least thrice integrable. Following the standard approach (e.g. [Ostrovsky & Potapov, 1999]), the equation (A.1) is integrated once, what gives

$$-Vu + 3|u|u + u_{\xi\xi} = C_\pm, \tag{A.2}$$

where $C_+$ and $C_-$ are the integration constants in the domains $u > 0$ and $u < 0$ respectively. Under the request that the second derivative $u_{\xi\xi}$ is continuous in the point of linkage $u(\xi_0) = 0$, Eq. (A.2) gives $C_+ = C_-$. For decaying at infinity $|x| \to \infty$ solutions, what in a physical problem corresponds to waves with finite energy, the parameters $C_+ = C_-$ should be chosen equal to zero. Therefore, we put $C_+ = C_- = 0$ in what follows. With this condition, if $u(\xi)$ is a solution of (A.2), then the function $-u(\xi)$ is the solution too, and then $u''(\xi_0) = 0$.

After multiplication of (A.2) by $u_\xi$ and performing the second integration, the equation yields

$$\frac{1}{2}u_\xi^2 + \Pi(u) = W_\pm, \quad \Pi(u) = |u|u^2 - \frac{V}{2}u^2, \tag{A.3}$$

where $W_\pm$ are the new integration constants (cf. Eq. (4)). Similar to the discussion just above, the continuity condition for the first derivative $u'(\xi_0)$ requires $W_+ = W_- = W$.

Comparing the potential $\Pi(u)$ in (A.3) with the potential for the classic KdV equation $P(v)$ in (4), we conclude that $\Pi(u) = P(u)$ for $u \geq 0$ and $\Pi(u) = P(-u)$ for $u < 0$ when putting in (4) $C = 0$. It is easy to see from the form of $P$, that these conditions require that the potential $\Pi(u)$ is composed from the part $P(v)$ for positive $v$ with its mirror-reflection, when the point $v = 0$ corresponds to one of local extremes of the function $P(v)$, see Fig. 1b,c.

Since the functions $u$ and $u'$ are continuous in the linkage point $\xi_0$ where $u$ changes the sign, the third derivative $u'''(\xi_0)$ is continuous too due to (A.1). However, for the fourth derivative the Eq. (A.1) gives:

$$u_{4\xi}(\xi_0) = \begin{cases} -(u_\xi(\xi_0))^2, & u > 0 \\ (u_\xi(\xi_0))^2, & u < 0 \end{cases}, \tag{A.4}$$

and according to (A.3),

$$(u_\xi(\xi_0))^2 = 2W. \tag{A.5}$$

The solution with $W = 0$ and thus with the zero first derivative $u'(\xi_0)$ can occur as a limiting case in the situation shown in Fig. 1c when the solution tends to a soliton. In all other situations of the solutions represented in Fig. 1b,c we have $W > 0$, and thus the fourth derivative experiences a jump in the points where the function $u(\xi)$ changes the sign.

## Appendix B. Nonlinear equation for long modulations of quasi-sinusoidal waves within the modular KdV equation

In this section we derive the equation on long modulations of weakly nonlinear waves governed by the modular KdV equation. For generality, in this section we use the equation with free real constants $\alpha$ and $\beta$ in the form

$$u_t + \alpha |u| u_x + \beta u_{xxx} = 0. \tag{B.1}$$

The equation (1) follows from (B.1) when one puts $\alpha = 6$ and $\beta = 1$. Let us introduce the complex envelope $\psi(x,t)$ for waves with some chosen carrier wavenumber $k_0$ and frequency $\omega_0$, which is a slow function of its variables $x$ and $t$:

$$u(x,t) = \frac{1}{2}\psi(x,t)e^{i\theta} + c.c., \quad \text{where} \quad \theta(x,t) = \omega_0 t - k_0 x. \tag{B.2}$$

The phase $\theta(x,t)$ is a fast function of its variables. The representation (B.2) is efficient only when the waves are close to monochromatic, i.e. the similarity parameter $Ur = a/k_0$, where $a$ is the characteristic wave amplitude, is not large. Otherwise, other wave harmonics should be taken into account.

We substitute the representation (B.2) to (B.1) and obtain by the direct calculations:

$$\psi_t e^{i\theta} + i\omega_0 \psi e^{i\theta} + \frac{\alpha}{2}\left|\psi e^{i\theta} + \psi^* e^{-i\theta}\right|\left(\psi_x e^{i\theta} - ik_0 \psi e^{i\theta}\right) +$$
$$+ \beta\left(ik_0^3 \psi e^{i\theta} - 3k_0^2 \psi_x e^{i\theta} - 3ik_0 \psi_{xx} e^{i\theta} + \psi_{xxx} e^{i\theta}\right) + c.c. = 0. \tag{B.3}$$

Neglecting higher-order dispersive corrections to the linear and nonlinear terms, we reduce the equation (B.3) to an approximate form

$$\psi_t e^{i\theta} + i\omega_0 \psi e^{i\theta} + \frac{\alpha}{2}\left|\psi e^{i\theta} + \psi^* e^{-i\theta}\right|\left(-ik_0 \psi e^{i\theta}\right) +$$
$$+ \beta\left(ik_0^3 \psi e^{i\theta} - 3k_0^2 \psi_x e^{i\theta} - 3ik_0 \psi_{xx} e^{i\theta}\right) + c.c. = O(\varepsilon^2 \mu) + O(\varepsilon \mu^3). \tag{B.4}$$

Here, small parameters of nonlinearity and dispersion are introduced for clarity, $\psi = O(\varepsilon)$ and $k_0^{-1}\partial_x = O(\mu)$, $\varepsilon \ll 1$, $\mu \ll 1$. The transformation from (B.3) to (B.4) implies the balance between small parameters $\varepsilon \sim \mu^2$, which form the new similarity parameter $\varepsilon/\mu^2 \sim a/K^2$, where $a$ is the wave amplitude and $K$ is the characteristic length of the modulation.

Further, we impose the dispersion relation

$$\omega_0 = -\beta k_0^3, \tag{B.5}$$

which follows from the linearized equation (B.1), and also introduce the group velocity:

$$c_{gr} = \frac{d\omega_0(k_0)}{dk_0} = -3\beta k_0^2. \tag{B.6}$$

Then the equation (B.4) reduces to the form

$$\left(\psi_t + c_{gr}\psi_x - 3ik_0\beta\psi_{xx}\right)e^{i\theta} - ik_0\frac{\alpha}{2}\left|\psi e^{i\theta} + \psi^* e^{-i\theta}\right|\psi e^{i\theta} + c.c. \approx 0. \tag{B.7}$$

We multiply the terms in (B.7) by $e^{-i\theta}$ and then integrate along the coordinate $x$ within one length of the dominant wave $L_0 = 2\pi/k_0$. Relying on the weak dependence of $\psi$ on $x$, this function and its derivatives are approximately constants within the interval of integration, therefore (B.7) yields:

$$2\pi(\psi_t + c_g\psi_x - 3ik_0\beta\psi_{xx}) - ik_0\frac{\alpha}{2}\int_{\theta_0-\pi}^{\theta_0+\pi}|\psi e^{i\theta} + \psi^* e^{-i\theta}|(\psi - \psi^* e^{-2i\theta})d\theta \approx 0. \quad (B.8)$$

For clarity, the period of integration along $x$ is converted to the integration by the phase $\theta$ in the interval around some chosen phase $\theta_0$ for the given instant of time. Note that other complex conjugate terms vanish as they give zeros after the integration.

It is now convenient to represent the complex amplitude in the form $\psi(x,t) = \rho(x,t)\exp(i\varphi(x,t))$, where $\rho$ and $\varphi$ are real functions, and to use the slowness of these functions with respect to the phase $\theta$:

$$2\pi(\psi_t + c_g\psi_x - 3ik_0\beta\psi_{xx}) - ik_0\frac{\alpha}{2}|\psi|\psi\int_{\theta_0-\pi}^{\theta_0+\pi}|e^{i(\theta+\varphi)} + e^{-i(\theta+\varphi)}|(1 - e^{-2i(\theta+\varphi)})d\theta \approx 0. \quad (B.9)$$

To proceed further, we use the decomposition into the Fourier series of the function $|\cos(\theta + \varphi)| = |\cos(k_0(x - x_0))|$, where for any given instant of time the reference coordinate $x_0(t)$ of the maximum of the function cosine is specified by the condition $k_0 x_0 = \omega_0 t + \varphi_0$. The decomposition may be obtained by the direct calculation:

$$|\cos(k_0(x - x_0))| = \frac{g_0}{2} + \sum_{n=1}^{\infty} g_n \cos(2k_0 n(x - x_0)), \quad g_n = \frac{4}{\pi}\frac{(-1)^{n+1}}{(4n^2 - 1)}. \quad (B.10)$$

The Fourier coefficients $g_n$ have alternating signs; by the absolute value they decay as $n^{-2}$. When the expansion (B.10) is substituted to the integral in (B.9), after the integration within the wave period, only two terms contribute to the value of the integral, and then the equation on $\psi(x,t)$ in a closed form appears:

$$\psi_t + c_g\psi_x - 3ik_0\beta\psi_{xx} - ik_0\alpha\left(\frac{g_0 - g_1}{2}\right)|\psi|\psi \approx 0. \quad (B.11)$$

It reduces to the form (19) when one takes the Fourier coefficients $g_n$ from (B.10) and puts $\alpha = 6$ and $\beta = 1$.

## Appendix C. Stability analysis of a perturbed plane wave within the modular NLS equation

In this section we analyze stability of regular waves within the envelope equation (19) derived in Appendix B. The perturbed plane wave (21) is represented in the form

$$\psi(x,t) = a_0(1 + \delta\chi(x,t))\exp(ipa_0 t), \quad (C.1)$$

where with no loss of generality we assume that the amplitude is positive, $a_0 > 0$. The perturbation $\chi(x,t)$ is complex-valued, and the modulation amplitude $\delta$ is a small parameter, $\delta \ll 1$. The ansatz (C.1) is substituted to the equation (19), and then the equation on the wave modulation is obtained in the first order of expansions for the small parameter $O(\delta)$:

$$i\chi_t + q\chi_{xx} + p\frac{a_0}{2}(\chi + \chi^*) = 0. \quad (C.2)$$

To obtain this relation, we have expanded the function $|\psi(x,t)|$, which is assumed to be close to the value $a_0$, using the Taylor series:

$$|\psi| = \left(|\psi|^2\right)^{\frac{1}{2}} = a_0\sqrt{1+\delta(\chi+\chi^*)+\delta^2|\chi|^2} = a_0\left(1+\frac{\delta}{2}(\chi+\chi^*)\right)+O(\delta^2). \quad \text{(C.3)}$$

We seek for the solution of the linear equation (C.2) in the form of a perturbation with the real wavenumber $K$ and the complex frequency $\Omega$:

$$\chi(x,t) = C_1 e^{i\Omega t - iKx} + C_2 e^{-i\Omega^* t + iKx}, \quad \text{(C.4)}$$

where $C_1$ and $C_2$ are constants. If the real and imaginary parts of the frequency are written explicitly, $\Omega = \Omega_R + i\Omega_I$, $\Omega_R \in \mathfrak{R}$, $\Omega_I \in \mathfrak{R}$, then (C.4) gives

$$\chi(x,t) = e^{-\Omega_I t}\left(C_1 e^{i\Omega_R t - iKx} + C_2 e^{-i\Omega_R t + iKx}\right), \quad \text{(C.5)}$$

and negative $\Omega_I$ will correspond to an exponential growth. From (C.4) and (C.2) the homogeneous system of algebraic equations on $C_1$ and $C_2$ may be obtained, with the compatibility condition

$$\Omega^2 = q^2 K^2 \left(K^2 - \frac{p}{q} A_0\right). \quad \text{(C.6)}$$

Therefore, the perturbations are unstable when

$$|K| < K_{BF}, \quad K_{BF} = \sqrt{\frac{p}{q} A_0}, \quad \text{(C.7)}$$

what requires $pq > 0$. The instability growth rate $\sigma = |\Omega_I|$ is maximum for the parameters

$$K_{max} = \sqrt{\frac{p}{2q} A_0}, \quad \sigma_{max} = \frac{\alpha A_0}{2}. \quad \text{(C.8)}$$

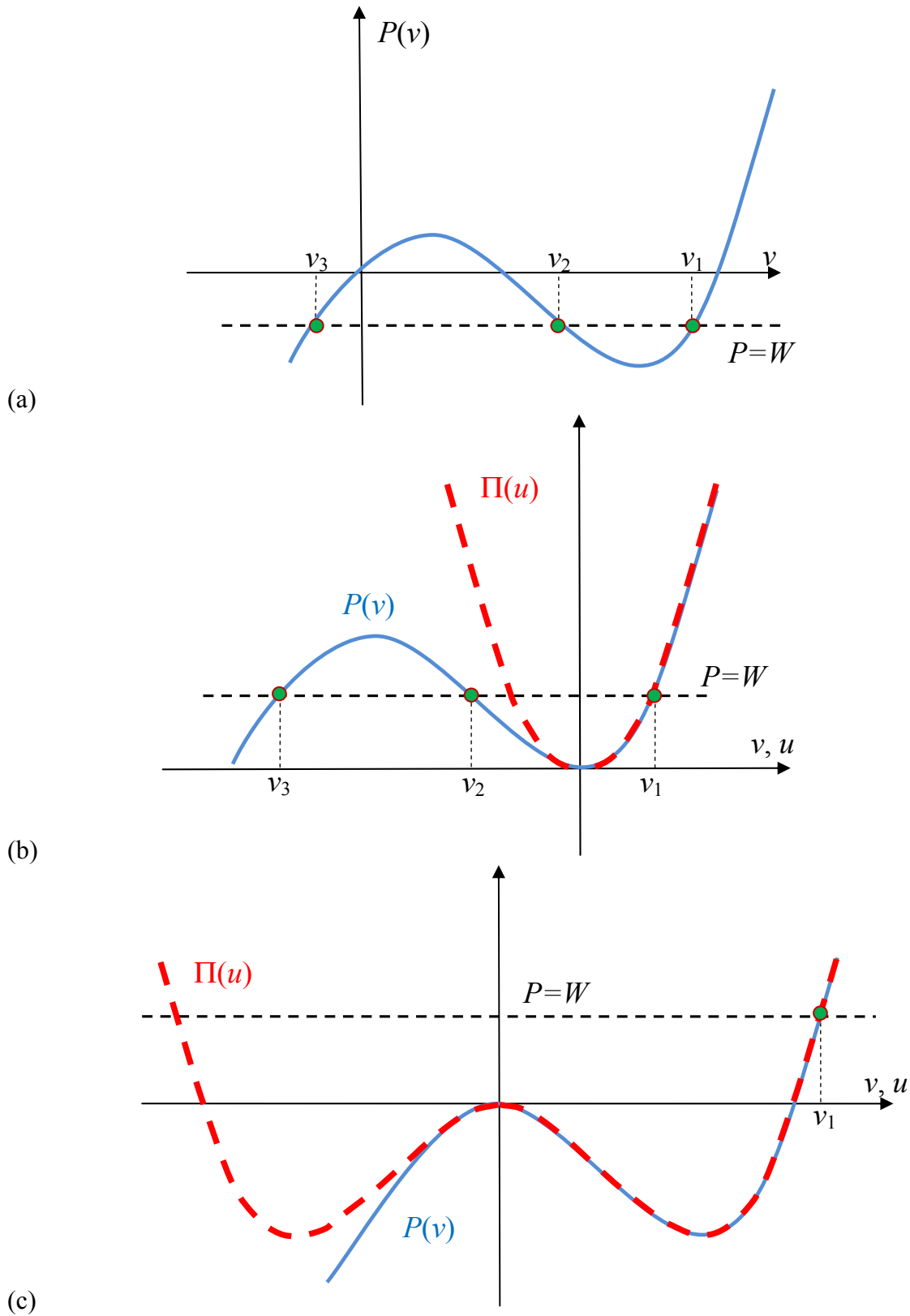

**Fig. 1.** Associated potential energies for the travelling solutions of the classic KdV equation (a) and the modular KdV equation with the linkage point at the local minimum (b) and local maximum (c). The functions $P(v)$ and $\Pi(u)$ is given by the blue solid line and dashed red line respectively. The horizontal dashed line denotes the level $W$.

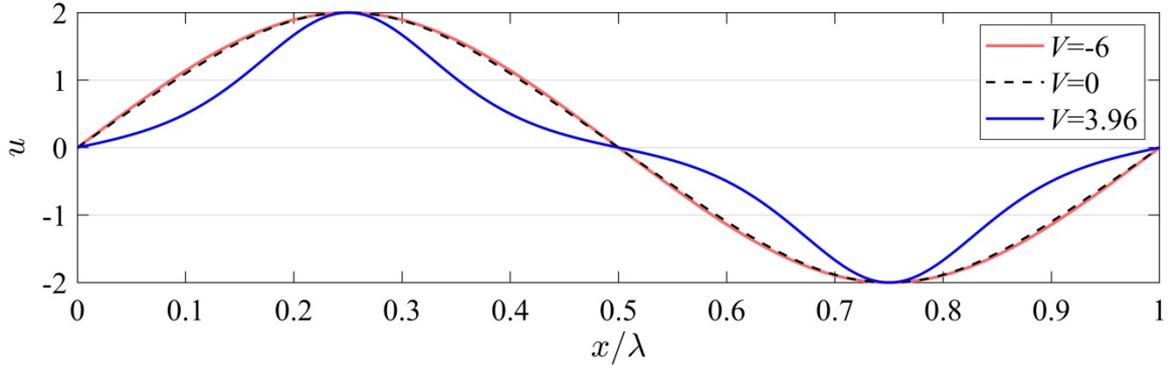

**Fig. 2.** Examples of nonlinear periodic waves for the modular KdV equation with the amplitude $a = 2$ and velocities $V = -6, 0, 3.96$ (see the legend). The corresponding wavenumbers are $k \approx 3.33, 2.24, 0.74$, and the Ursell numbers are $a/k^2 \approx 0.18, 0.40, 3.70$.

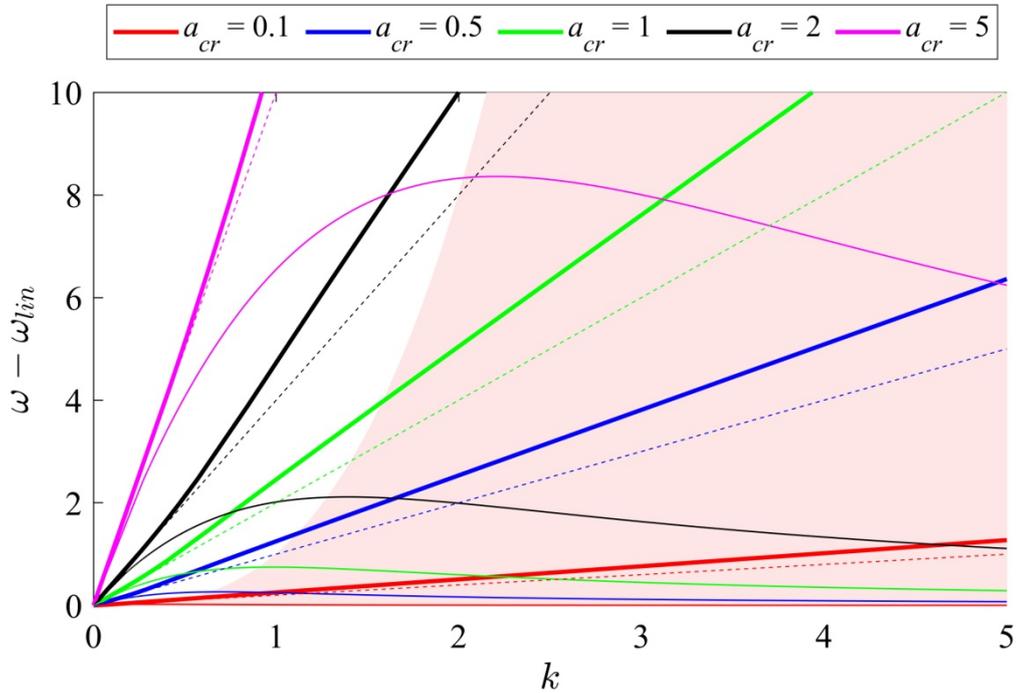

**Fig. 3.** Nonlinear dispersion relation for the modular KdV equation (thick solid curves) compared with the relation for the classic KdV equation (thin solid curves). The soliton limits for $k \to 0$ are shown with dotted straight lines. The curve colors correspond to the same choices of the crest amplitudes $a_{cr}$, see the legend. The shaded area corresponds to waves with negative velocities $V < 0$.

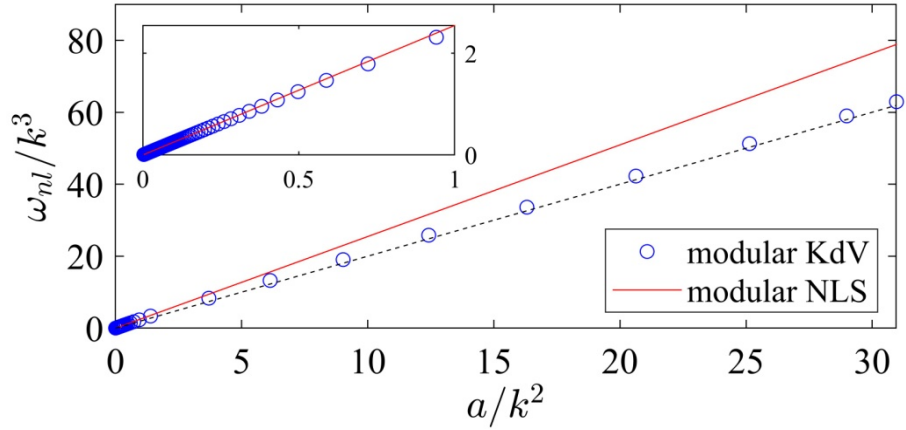

**Fig. 4.** Nonlinear frequency shift of the plane wave solution of the modular NLS equation (21) (solid line) and of the nonlinear periodic waves of the modular KdV equation (symbols) with $k = k_0$ and $a = a_0$. The dashed straight line corresponds to the soliton limit of very long waves according to the modular KdV equation.

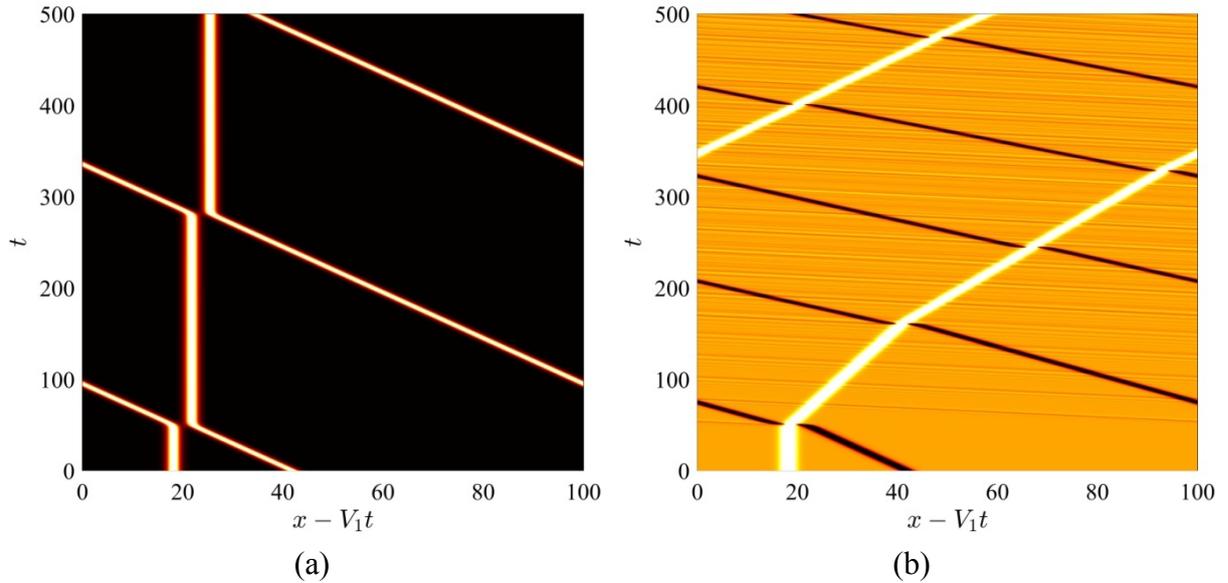

(a)          (b)

**Fig. 5.** Interactions of two KdV solitons within the modular equation with amplitudes of the same sign, $A_1 = 2$ and $A_2 = 1.8$ (a), and of different signs, $A_1 = 2$ and $A_2 = -1.8$ (b). The reference velocity corresponds to the biggest soliton speed, $V_1 = 2A_1$. The pseudocolor in panel (b) is artificially contrasted to make the radiated waves more visible.

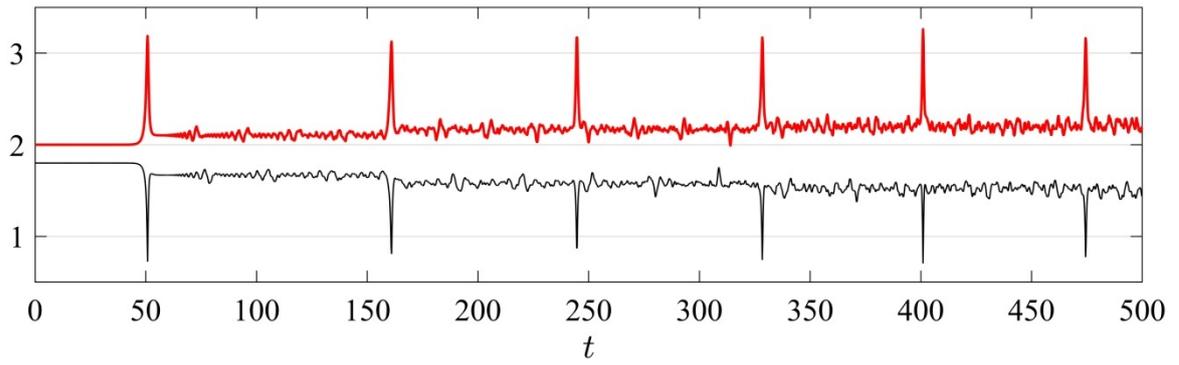

**Fig. 6.** The maximum $\max_x(u)$ and the minimum $-\min_x(u)$ as functions of time (the upper and lower curves respectively) for the numerical experiment shown in Fig. 5b.

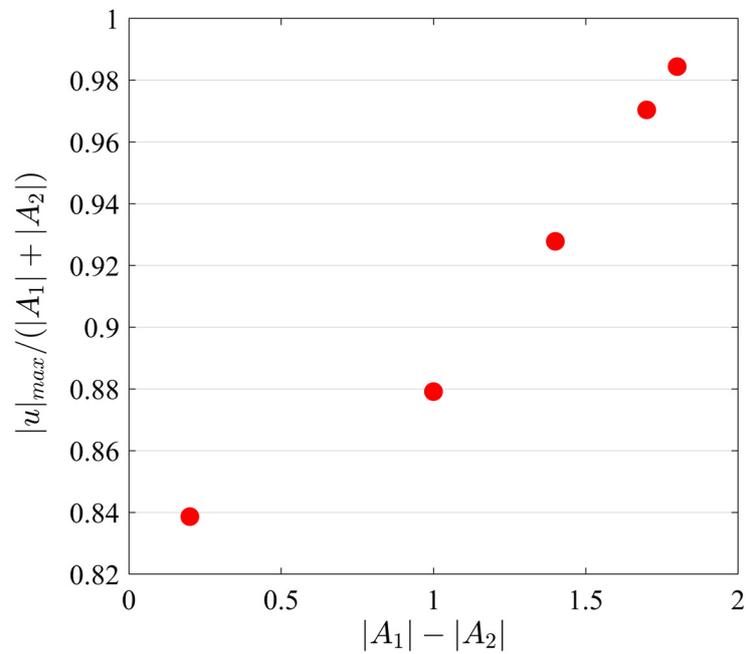

**Fig. 7.** Maximum value of the solution $|u|_{max}$ attained in the course of interaction between two solitons of different signs within the modular KdV versus the ideal superposition $|A_1|+|A_2|$. The soliton amplitudes are $A_1 = 2$ and $-2 < A_2 < 0$.

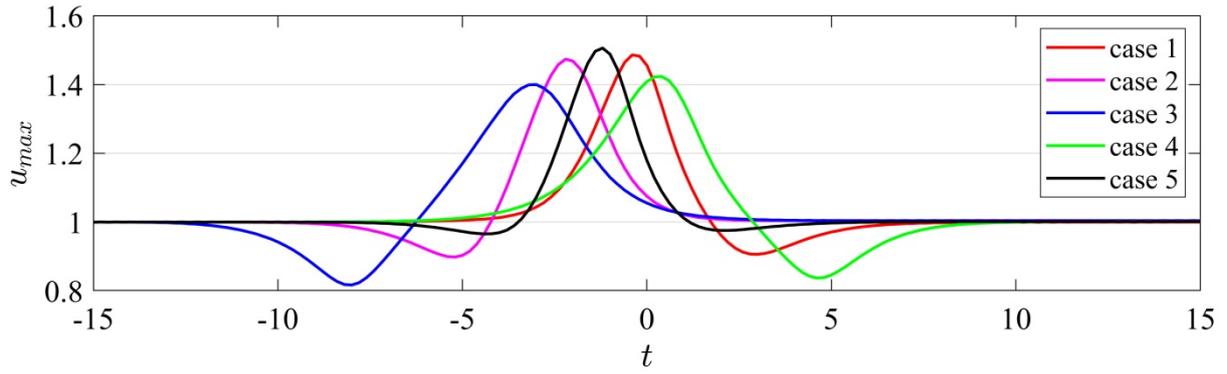

**Fig. 8.** Evolution of the wave maxima $\max_x(u)$ in time in simultaneous interactions between solitons with amplitudes $A_1 = 1$, $A_2 = -0.6$ and $A_3 = 0.2$ (similar to the simulations shown in Fig. 9). Different curves correspond to slightly different initial position of the third soliton.

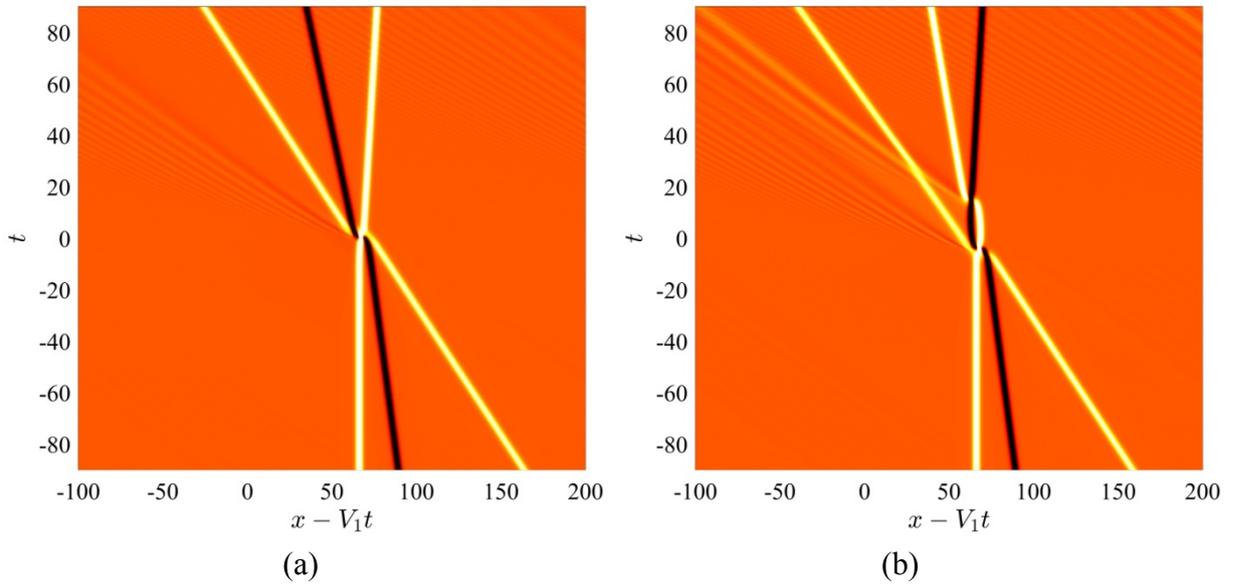

**Fig. 9.** Examples of interaction between three solitons of alternating signs $A_1 = 1$, $A_2 = -0.9$, $A_3 = 0.5$ within the modular KdV equation. The initial conditions in (a) and (b) differ in the original location of the smallest soliton.

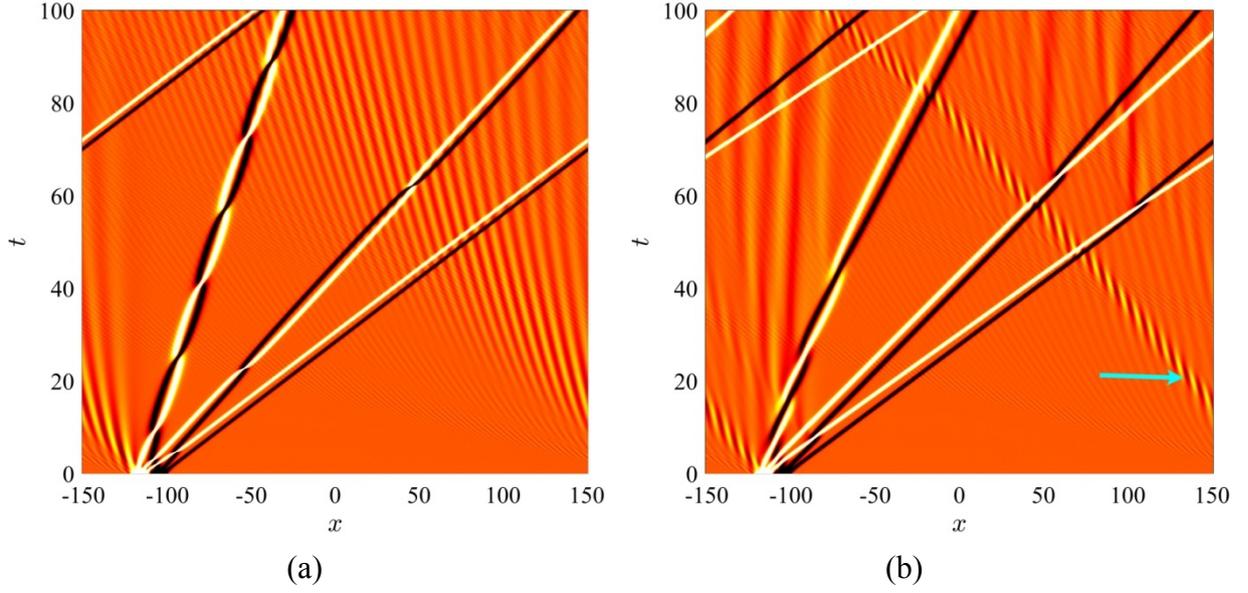

**Fig. 10.** Numerical simulations of the initial-value problem for the same localized sign-changing perturbation within the modified KdV equation (a) and within the modular KdV equation (b). The arrow in panel (b) points at the stable breather.

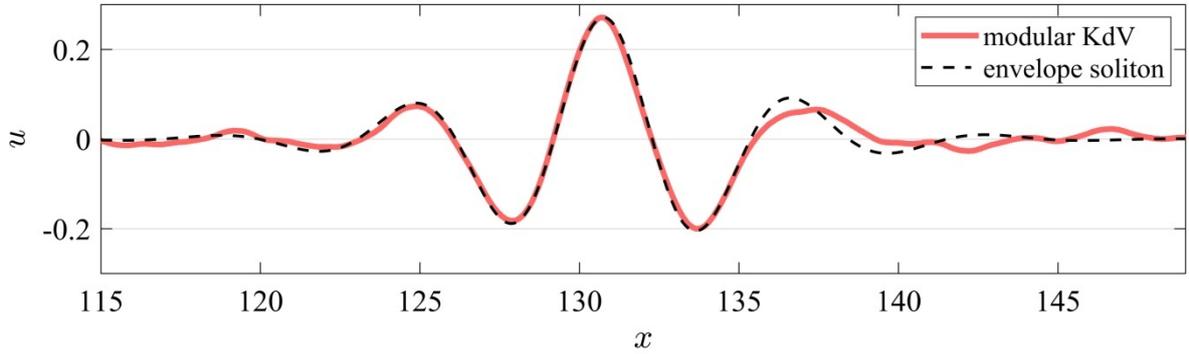

**Fig. 11.** Fitting the envelope soliton solution (22) to the isolated wave group in the numerical simulation shown in Fig. 10b at $t = 23.2$ (see the arrow). The fitting parameters are $k_0 = 1$, $x_0 = 131$, $B = 0.272$.

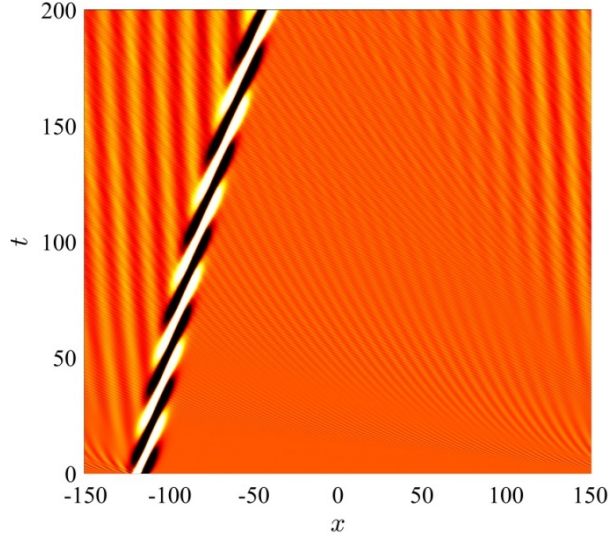

**Fig. 12.** Numerical simulation of the initial-value problem for a sign-changing perturbation within the modular KdV, when a stable breather propagating to the right emerges. The pseudocolor is artificially contrasted to make the radiated waves more visible.

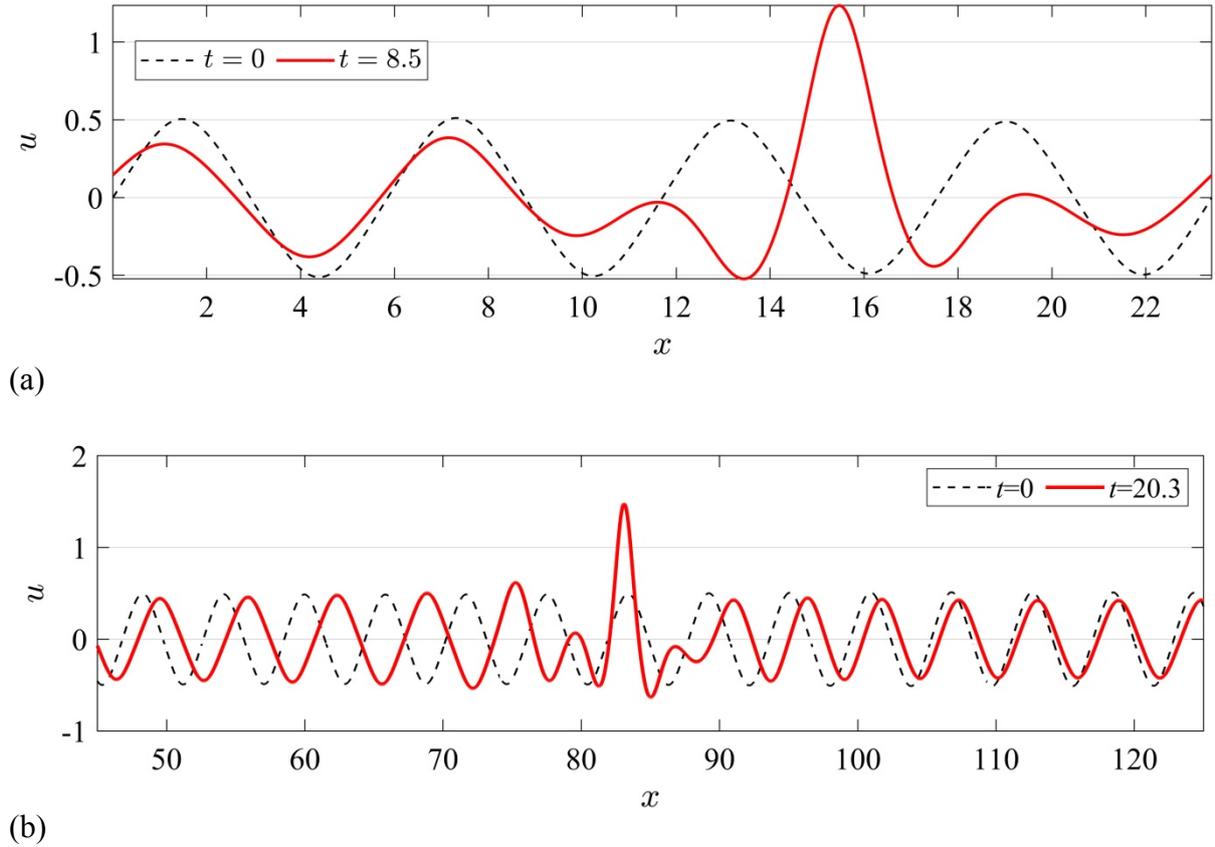

(a)

(b)

**Fig. 13.** Initial conditions (dashed lines) and the maximum waves (solid lines) occurred as a result of the modulational instability of weakly perturbed waves within the modular KdV equation. The initial wave parameters are $a_0 = 0.5$, $k_0 \approx 1.07$, $a_0/k_0^2 \approx 0.43$. The perturbation lengths are equal to $N_w = 4$ waves (a) and $N_w = 15$ waves (b).

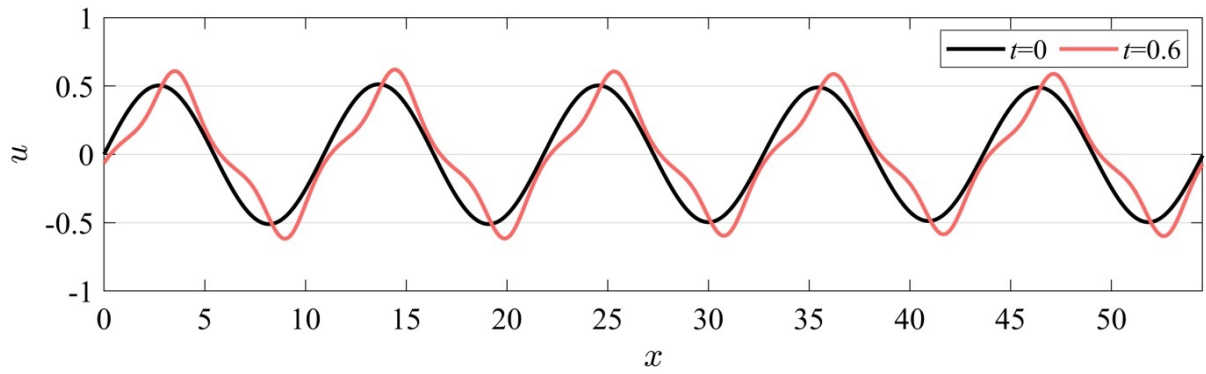

(a)

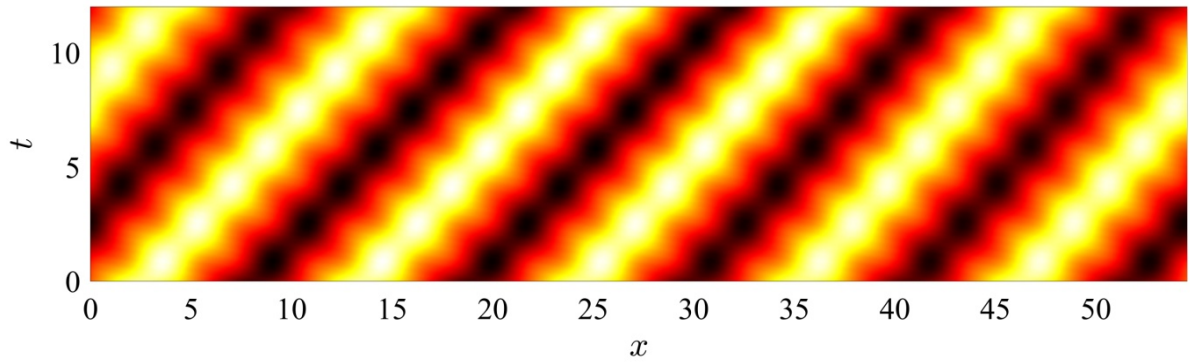

(b)

**Fig. 14.** Two snapshots (a) and the space-time diagram (b) of the solution $u(x,t)$ of the modular KdV equation for the initial condition in the form of a sinusoidal wave with $a = 0.5$ and $k \approx 0.58$, $a/k^2 \approx 1.5$.